\documentclass[]{aastex631}

\usepackage{hyperref}
\usepackage{multirow}

\shortauthors{Lu et al.}

\begin{document}

\title{Periodic Coronal Rain Driven by Self-consistent Heating Process in a Radiative Magnetohydrodynamic Simulation}

\correspondingauthor{Feng Chen}
\email{chenfeng@nju.edu.cn}

\author[0000-0001-7961-7617]{Zekun Lu}
\affiliation{School of Astronomy and Space Science, Nanjing University, Nanjing 210023, China}
\affiliation{Key Laboratory of Modern Astronomy and Astrophysics (Nanjing University), Ministry of Education, Nanjing 210023, China}

\author[0000-0002-1963-5319]{Feng Chen}
\affiliation{School of Astronomy and Space Science, Nanjing University, Nanjing 210023, China}
\affiliation{Key Laboratory of Modern Astronomy and Astrophysics (Nanjing University), Ministry of Education, Nanjing 210023, China}

\author[0000-0002-4205-5566]{J. H. Guo}
\affiliation{School of Astronomy and Space Science, Nanjing University, Nanjing 210023, China}
\affiliation{Key Laboratory of Modern Astronomy and Astrophysics (Nanjing University), Ministry of Education, Nanjing 210023, China}
\affiliation{Centre for mathematical Plasma Astrophysics, Department of Mathematics, KU Leuven, Celestijnenlaan 200B, B-3001 Leuven, Belgium}

\author[0000-0002-4978-4972]{M. D. Ding}
\affiliation{School of Astronomy and Space Science, Nanjing University, Nanjing 210023, China}
\affiliation{Key Laboratory of Modern Astronomy and Astrophysics (Nanjing University), Ministry of Education, Nanjing 210023, China}

\author[0000-0002-6799-4340]{Can Wang}
\affiliation{School of Astronomy and Space Science, Nanjing University, Nanjing 210023, China}
\affiliation{Key Laboratory of Modern Astronomy and Astrophysics (Nanjing University), Ministry of Education, Nanjing 210023, China}

\author[0009-0005-8311-1703]{Haocheng Yu}
\affiliation{School of Astronomy and Space Science, Nanjing University, Nanjing 210023, China}
\affiliation{Key Laboratory of Modern Astronomy and Astrophysics (Nanjing University), Ministry of Education, Nanjing 210023, China}

\author[0000-0002-9908-291X]{Y. W. Ni}
\affiliation{School of Astronomy and Space Science, Nanjing University, Nanjing 210023, China}
\affiliation{Key Laboratory of Modern Astronomy and Astrophysics (Nanjing University), Ministry of Education, Nanjing 210023, China}

\author[0000-0002-7153-4304]{Chun Xia}
\affiliation{School of Physics and Astronomy, Yunnan University, Kunming 650050, China}



\begin{abstract}

The periodic coronal rain and in-phase radiative intensity pulsations have been observed in multiple wavelengths in recent years. However, due to the lack of three-dimensional coronal magnetic fields and thermodynamic data in observations, it remains challenging to quantify the coronal heating rate that drives the mass cycles. In this work, based on the MURaM code, we conduct a three-dimensional radiative magnetohydrodynamic simulation spanning from the convective zone to the corona, where the solar atmosphere is heated self-consistently through dissipation resulting from magneto-convection. For the first time, we model the periodic coronal rain in an active region. With a high spatial resolution, the simulation well resembles the observational features across different extreme ultraviolet wavelengths. These include the realistic interweaving coronal loops, periodic coronal rain and periodic intensity pulsations, with two periods of 3.0~h and 3.7~h identified within one loop system. Moreover, the simulation allows for a detailed three-dimensional depiction of coronal rain on small scales, revealing adjacent shower-like rain clumps $\sim500$~km in width and showcasing their multi-thermal internal structures. We further reveal that these periodic variations essentially reflect the cyclic energy evolution of the coronal loop under thermal non-equilibrium state. Importantly, as the driver of the mass circulation, the self-consistent coronal heating rate is considerably complex in time and space, with hour-level variations in one order of magnitude, minute-level bursts, and varying asymmetry reaching ten times between footpoints. This provides an instructive template for the ad hoc heating function, and further enhances our understanding of the coronal heating process.

\end{abstract}

\keywords{Solar corona (1483) --- Solar coronal loops (1485) --- Solar extreme ultraviolet emission (1493) --- Solar magnetic fields (1503) --- Solar active regions (1974) --- Solar coronal heating (1989) --- Solar convective zone (1998) --- Radiative magnetohydrodynamics (2009)}

\section{Introduction} \label{sec:intro}

Coronal rain refers to elongated, cool and dense clumps forming in the corona and falling down to the solar surface, visible mostly in chromospheric and transition region spectral lines \citep{Schrijver-2001SoPh..198..325S,DeGroof-2005A&A...443..319D,Antolin-2012ApJ-fine-rains...745..152A,Jing-2016NatSR...624319J,LiLP-2018ApJ...864L...4L,Schad-2018ApJ-HeI...865...31S,Mason-2019ApJ-fanspine-rain...874L..33M,Chen-2022A&A-downflows...659A.107C,Antolin-2023A&A...676A.112A}. Moreover, in recent years, coronal rain has been observed to recur in a time period of hours \citep{Auchere-2018ApJ...853..176A,Froment-2020A&A...633A..11F,Sahin-2023ApJ-upward-rain...950..171S}, concurrent with the widespread long-period pulsations in extreme ultraviolet (EUV) intensity of the coronal loops \citep{Auchere-2014A&A...563A...8A,Auchere-2016ApJ...827..152A,Froment-2015ApJ-DEM...807..158F}.

According to the modern understanding, coronal rain results from catastrophic cooling within coronal loops under thermal non-equilibrium (TNE; \citealt{Antiochos-1991ApJ...378..372A,Klimchuk-2019SoPh-ti-tne..294..173K,Antolin-2020PPCF-review...62a4016A}), where the intense heating at footpoints evaporates plasma to fill the loop, but as density increases, the relatively weaker heating rate at the loop top cannot offset the growing local radiative losses. Such energy losses are rapidly amplified through thermal instability (TI; \citealt{Parker-1953ApJ...117..431P,Field-1965ApJ...142..531F,Claes-2020A&A...636A.112C,Claes-2021SoPh..296..143C}), causing the coronal plasma to cool down dramatically and form dense rain clumps. The state of TNE will persist as the heating rate remains temporally quasi-steady, leading to periodic evaporation and condensation of the coronal loop. 

Such a formation and evolutionary picture has been extensively studied through numerical simulations in different dimensions. One-dimensional (1D) parameterized hydrodynamic simulations focus on understanding the conditions contributing to TNE or the occurrence of periodic coronal rain, including aspects such as heating stratification \citep{Muller-2004A&A...424..289M,Downs-2016ApJ...832..180D}, asymmetry \citep{Froment-2018ApJ-1d-tne...855...52F,Klimchuck-2019ApJ-tne...884...68K}, timescales \citep{Johnston-2019A&A...625A.149J} and geometry of the hosting coronal loop \citep{Mikic-2013ApJ...773...94M,Pelouze-2022A&A...658A..71P}, which closely links coronal rain with the long baffling coronal heating problem \citep{Antolin-2010ApJ...716..154A}. 2.5D \citep{Fang-2013ApJ...771L..29F,Fang-2015ApJ...807..142F,Lixh-2022ApJ-rain...926..216L} and 3D \citep{Xia-2017A&A-rain...603A..42X} magnetohydrodynamics (MHD) simulations further revealed coronal rain's dynamics, fine structures and interactions with coronal magnetic fields. However, considering the episodic and time-varying nature of coronal heating rates \citep{Hansteen-2015ApJ...811..106H,Testa-2020ApJ...889..124T,ChenF-2022ApJ...937...91C,Lu-2024NatAs...8..706L}, the user-defined heating functions employed in these simulations remain overly simplified by comparison.

Making use of the Bifrost code \citep{Gudiksen-bifrost-2011A&A...531A.154G}, \cite{Kohutova-2020A&A...639A..20K} and \cite{Antolin-2022ApJ...926L..29A} achieved higher degrees of self-consistency and realism by attributing heating rates to sum of ohmic and viscous dissipation, ultimately driven by the magneto-convection. They modelled and investigated one-off coronal rains forming in the quiet Sun, while it lacks degrees of generality and demands further study whether the self-consistent setup can reproduce periodic coronal rains in active regions, where such phenomena are more commonly observed \citep{Antolin-2020PPCF-review...62a4016A}. 

In this Letter, we conduct a 3D radiative magneto-hydrodynamics (RMHD) simulation based on the MURaM code \citep{Vogler-2005A&A...429..335V,Rempel-2017ApJ...834...10R}, where the heating process is realized self-consistently by including the convective motions, and the magnetic field strength in the photosphere reaches over 2000~G. For the first time, we successfully simulate the formation and evolution of periodic coronal rains in an active region, with realistic synthetic observations reproduced from various aspects. As a result, we can derive 3D magnetic and thermodynamic properties of coronal rains, which further sheds light on the underlying coronal heating and cooling processes.

\section{Numerical Simulation Setup} \label{sec:method}

The numerical model is based on the active region setup (the AR case therein) in \cite{Rempel-2017ApJ...834...10R}. The AR simulation started by adding a pair of slightly asymmetric bipolar sunspots to a near-surface convective zone simulation of quiet Sun. It was then extended to corona by raising the top boundary and initializing with a potential magnetic field extrapolation and a hydrostatic isothermal atmosphere in the coronal domain. Such extension was first implemented to 8 Mm and then further to around 41 Mm above the photosphere, with subsequent relaxation following each stage. Plasma is heated self-consistently through the sum of numerical viscous heating and resistive heating, where the two terms measure the dissipation of kinetic energy and magnetic energy, respectively. The bottom boundary condition is symmetric for magnetic field components and outflow mass flux, with the mean gas pressure kept fixed and its perturbations damped; the entropy of inflows is specified such that the stratification is inline with the standard solar model. A comprehensive description of the boundary condition (``O16b'' therein) can be found in \cite{Rempel-2014ApJ...789..132R}.

In this work, we first adopt a snapshot of the AR simulation from \cite{Rempel-2017ApJ...834...10R}, which corresponds to a state when the thermodynamic properties of the coronal plasma have reached a dynamic equilibrium. Then, we refine the mesh with nearest neighbor interpolation from $192\times192\times64$~km$^{3}$ to $96\times96\times32$~km$^{3}$ as the new initial state. The new run is then evolved till the average coronal temperature and density reach a new dynamic equilibrium (Figure \ref{efig:relaxation}).

\begin{figure*}[htb]
\centering
\includegraphics[width=\textwidth]{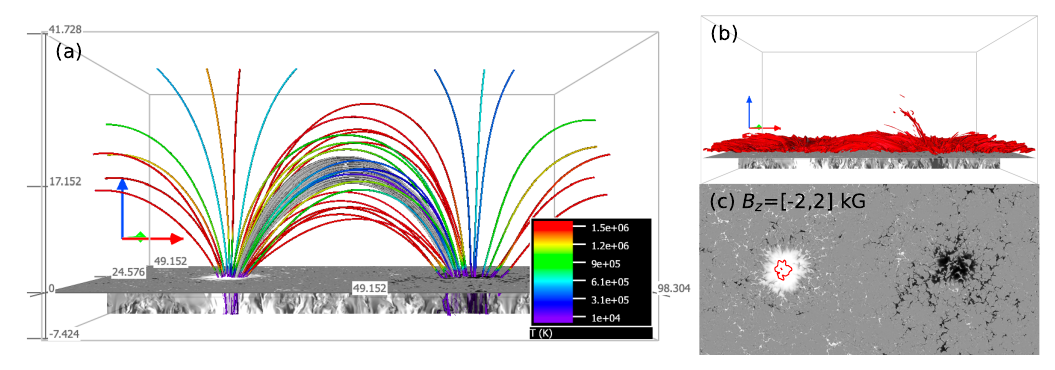} 
\caption{(a) A 3D overview of the simulation result at $t=$~2~h~18.0~min, where the colored tubes represent magnetic field lines, with colors indicating the temperature. The central grey tubes represent the magnetic field lines through the 91 fixed seed points, which are used for quantitative calculations in Figures \ref{fig4:tne} and \ref{fig5:heating_rates}. These points are chosen with positions as follows: $x$ equals 49.104 Mm; $y$ ranges from 22 to 25~Mm, with a 0.5~Mm spacing; $z$ ranges from 18.5 to 24.5~Mm, with a 0.5~Mm spacing. The grey-scale slices at the bottom display the $z-$direction magnetic field component, $B_z$, in the photosphere and of the convective zone.  The red, green, and blue arrows denote the $x-$, $y-$, and $z-$axes, respectively. (b) The number density contour of $5\times10^9\ \rm{cm^{-3}}$. (c) The magnetogram of the photosphere, where the colors from black to white represent values ranging from -2 to 2~kG. The contour of $B_z = 2.5\ \rm{kG}$ is overlaid in red.}
\label{fig:overview}
\end{figure*}

The analyses in this work start from the time instant of $t=1\ \rm{h}$. An overview of the numerical model is shown in Figure~\ref{fig:overview}a, where the horizontal domain covers an area of $98.304\times49.152$ Mm$^{2}$ and the vertical domain extends to 49.152 Mm, including a convective zone with a depth of 7.424 Mm below the photosphere. The active region maintains the initial asymmetry, with the positive polarity more coherent and over $500\ \rm{G}$ stronger than the negative polarity, as indicated by the photosphere magnetogram (Figure~\ref{fig:overview}c).

\begin{figure*}
\centering
\includegraphics[width=1.0\textwidth]{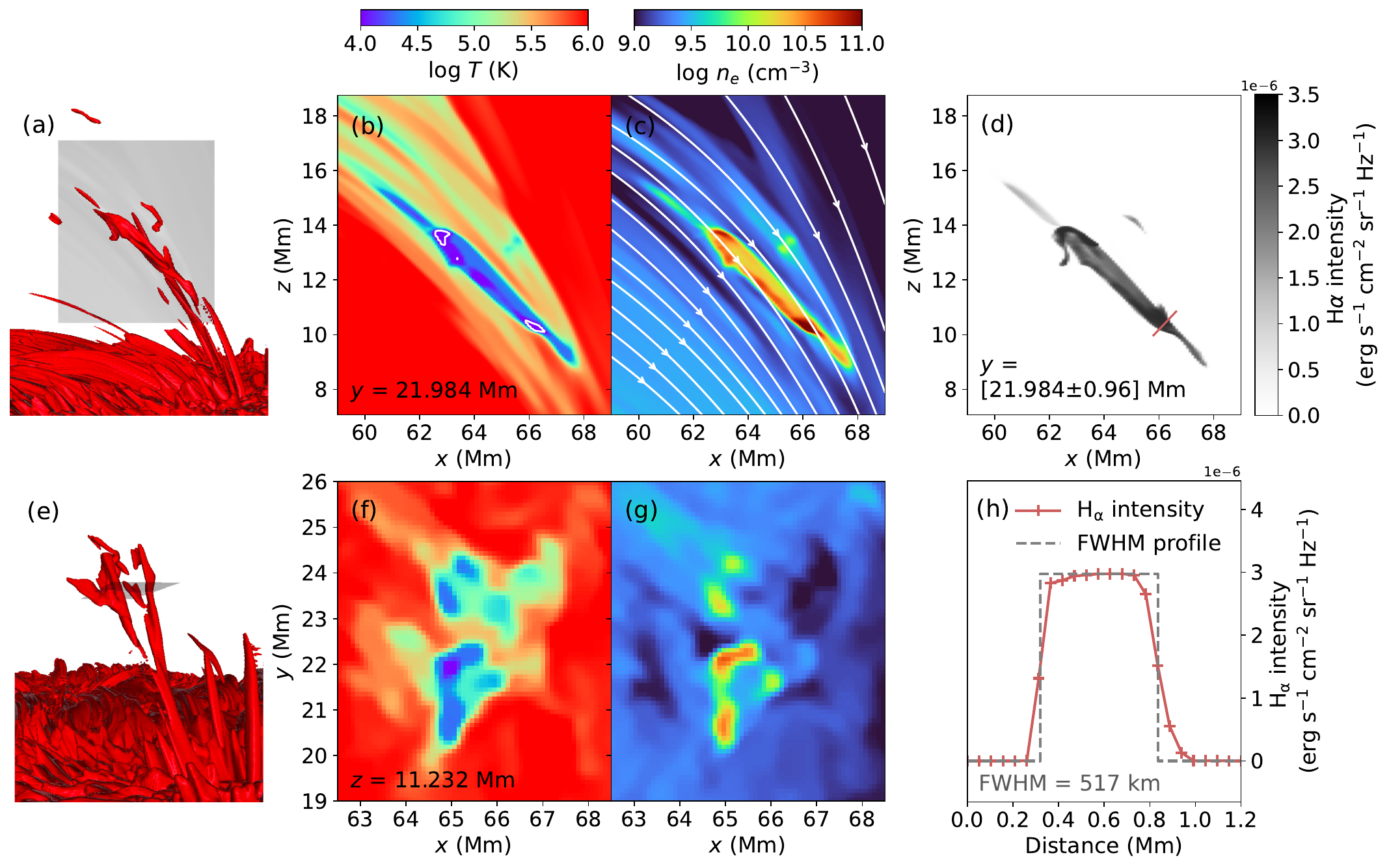} 
\caption{(a) A zoom-in view of the number density contour in Figure~\ref{fig:overview}(b). The grey slice crosses the coronal rain in the $x-z$ plane. (b) 2D distribution of temperature within this plane, where the white contour denotes the temperature of $T=8000$~K. (c) 2D distribution of number density, with white streamlines representing magnetic field lines projected on the plane. (d) 2D distribution of synthetic $\mathrm{H}\alpha$ intensity, with the radiative transfer considered across a distance of 1.92~Mm along the $y$-axis. Similarly, panels (e)-(g) present the same quantities but with the 2D slice in the $x-y$ plane. (h) The $\mathrm{H}\alpha$ intensity measured along the red slit in panel (d), where the grey dashed line represents its Full Width at Half Maximum (FWHM) profile.
}
\label{fig2:rain_slcie}
\end{figure*}

\section{Results} \label{sec:results}

\begin{figure*}
\centering
\includegraphics[width=\textwidth]{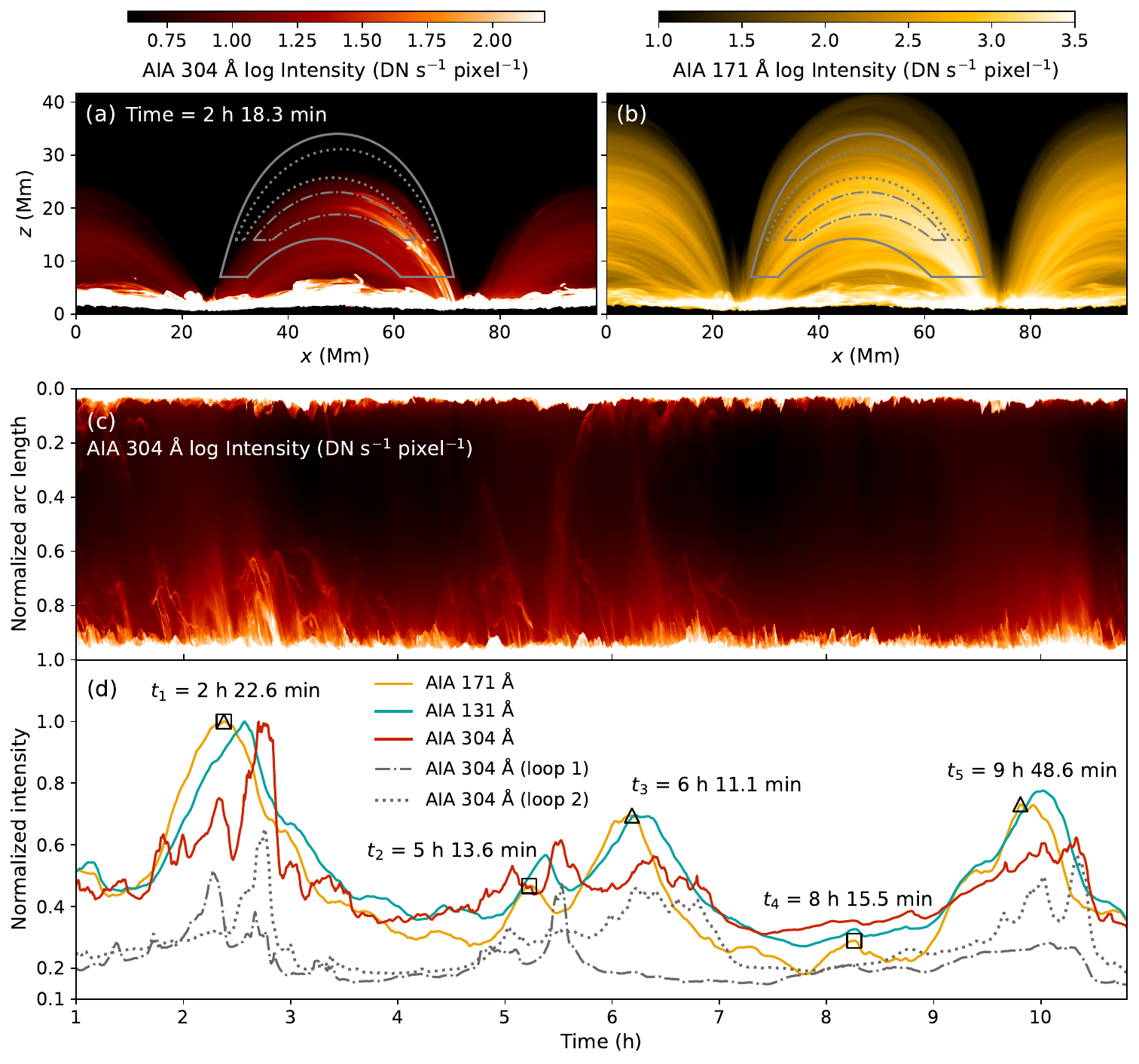} 
\caption{Synthetic radiative intensities integrated along the $y-$direction in (a) AIA $304\ \rm{\AA}$ and (b) $171\ \rm{\AA}$, where the solid grey line encloses the integration region for solid light curves of the coronal loop system in panel (d). The dash-dot and dotted lines enclose the integration regions for the light curves of coronal loop 1 and 2 in panel (d), respectively. (c) Temporal evolution of the synthetic AIA $304\ \rm{\AA}$ intensity along the loop system, based on averaging along the bundle of 100 magnetic field lines as outlined by the grey solid arcs in panel (a). The normalized arc length in panel (c) begins at the reference layer of $z=1.024$~Mm and ranges from 0 to 1, representing the path from the magnetic positive to negative polarity, along the increasing $x-$axis direction. (d) Normalized light curves of AIA $171\ \rm{\AA}$, $131\ \rm{\AA}$ and $304\ \rm{\AA}$. The light curves of loops 1 and 2 are normalized by the maximum $304\ \rm{\AA}$ intensity of the large enclosed region, but are magnified 2.8 and 4.2 times, respectively, aiming to provide a clearer visual contrast. The squares and triangles mark the peaks of the $171\ \rm{\AA}$ light curve of loops 1 and 2, respectively. The peak times are recorded and analyzed in Table \ref{tab:period}. An animation of this figure is available online, which depicts evolution of the synthetic radiative intensities in panels (a) and (b) from $t=1$~h to $t=10$~h~48~min, with a gray dashed line in panels (c) and (d) marking the time instant.
}
\label{fig3:euv_aia}
\end{figure*}

\subsection{Characteristics of coronal rain} \label{subsec:the rain}

The convective motions in the lower layer continuously drive evolution of magnetic fields, further leading to the thermodynamic evolution of the plasma, including cooling and condensations. Importantly, in addition to the optically-thin radiative loss calculated with CHIANTI database \citep{Landi-2012ApJ...744...99L}, this simulation also considers the 3D grey radiative transfer using short characteristics method \citep{Vogler-2005A&A...429..335V}. Such optically-thick radiation applies to the plasma with temperatures $T<2\times10^4$~K and optical depth $\tau>10^{-8}$ (the opacity and source function are set zero otherwise). This ensures a more accurate coronal cooling process. For instance, at $t=$~2~h~18.0~min, a group of coronal rains form at the leg region near the negative polarity (Figures \ref{fig:overview}b and \ref{fig2:rain_slcie}a). The coronal rain exhibits significant inhomogeneities in both temperature and number density distributions (Figure~\ref{fig2:rain_slcie}b,~c), characterized by three main components: (1) the thin and sharp ``envelope'' with transition region temperatures of $T\sim10^5$~K, which well corresponds to the condensation–corona transition region (CCTR; \citealt{Fang-2015ApJ...807..142F,Antolin-2022ApJ...926L..29A}), (2) the main ``body'' extending over around $7$~Mm with temperatures in the lower transition region of $T\lesssim3\times10^4$~K, and (3) several small-scale, cold and dense ``cores'' with chromospheric temperatures of $T<10^4$~K, where the cores located at the head and tail of the rain even cool down to below 8000~K. This aligns with the multi-thermal features of coronal rain inferred from different spectra in observations \citep{Antolin-2015ApJ...806...81A}. Moreover, our simulation results provide further references for understanding its internal temperature and density structures in three dimensions.

Since the observed emission arises from the collective effect of temperature and number density, coupled with the radiative transfer process including absorption, to compare with observations, we synthesize the widely-used $\rm{H}\alpha$ line center emission (see appendix for details) and obtain a realistic inhomogeneous profile (Figure~\ref{fig2:rain_slcie}d). The strongest emission concentrates on the head and tail. We measure the width of the rain head as FWHM~$\sim$~517~km (Figure~\ref{fig2:rain_slcie}h), falling within the observational statistics of Swedish 1-m Solar Telescope (SST; \citealt{Scharmer-2003SPIE.4853..341S}) while on the larger side of the statistical average \citep{Antolin-2012ApJ-fine-rains...745..152A,Froment-2020A&A...633A..11F}.

Moreover, by viewing the coronal rain from another angle (Figure~\ref{fig2:rain_slcie}e-g), one can find another significant spatial feature that a group of rain clumps are falling simultaneously within a neighbouring region of $\sim5$~Mm, which well reproduces the rain ``showers'' \citep{Xia-2017A&A-rain...603A..42X} that are identified in $\rm{H \alpha}$ \citep{Antolin-2012ApJ-fine-rains...745..152A,Froment-2020A&A...633A..11F} and EUV \citep{Antolin-2023A&A...676A.112A} images. This indicates that these adjacent coronal strands are experiencing common heating and cooling processes, which will be further investigated from an energy perspective in Section \ref{subsec:rain and EUV pulsation}.

\subsection{Periodic coronal rains} \label{subsec:rain and EUV pulsation}

To demonstrate the large-scale and long-term radiative evolution of coronal loops in the active region, we synthesize Atmospheric Imaging Assembly (AIA; \citealt{Lemen-2012SoPh..275...17L}) EUV emissions based on the optically thin scheme \citep{Boerner-2012SoPh..275...41B}. The AIA $171\ \mathrm{\AA}$ image exhibits a system of abundant coronal loops with numerous interweaving fine structures (Figure~\ref{fig3:euv_aia}b), with the rain ``shower'' clearly visible in $304\ \mathrm{\AA}$ (Figure~\ref{fig3:euv_aia}a). Figure~\ref{fig3:euv_aia}c presents the temporal evolution of AIA $304\ \mathrm{\AA}$ intensity along the loop system, which well replicates the two key features of the long-term observations of coronal rains \citep{Auchere-2018ApJ...853..176A, Froment-2020A&A...633A..11F}. Firstly, the coronal rains manifest as multiple shower-like brightenings in the time-distance diagram, indicating that the catastrophic cooling occurs simultaneously in neighboring coronal strands. Secondly, the coronal rain showers exhibit a prominent cyclic pattern, occurring intensely and then recurring twice with a time interval of several hours, much like a monsoon \citep{Auchere-2018ApJ...853..176A}.

To explore the periodic characteristics within the temporal evolution, we analyze EUV light curves of the coronal loop system (Figure~\ref{fig3:euv_aia}d) by three steps.

\begin{enumerate}
\itemsep0em
    \item We identify five prominent peaks ($t_1-t_5$) in the light curve at AIA~$\rm{171\ \AA}$, each followed by peaks at AIA~$\rm{131\ \AA}$ and $\rm{304\ \AA}$ in succession as observed in \cite{Auchere-2018ApJ...853..176A}. Such a temporal shift indicates a coronal cooling process with temperatures decreasing from around $0.9$~MK to $0.5$~MK and finally to around $0.09$~MK, accompanied with formation coronal rains (see the online animation of Figure~\ref{fig3:euv_aia}). Such a cooling feature is relatively not prominent for the weak peak of time $t_3$. 
    \item In terms of temporal evolution, we categorize the five peak moments into two groups: $t_1$, $t_2$, $t_4$, and $t_1$, $t_3$, $t_5$, with corresponding average periods of 2.94~h and 3.72~h (Table \ref{tab:period}), respectively.
    \item Regarding spatial aspects, by integrating the emission within two smaller arched regions as shown in Figure~\ref{fig3:euv_aia}a (outlined by dotted and dash-dot lines), we further ascribe the two groups of peaks to two different coronal loops.
\end{enumerate}
The findings above reveal a noteworthy observation: the overall cyclic lightcurves of the loop system arises from the superposition of periodic evolution of two different coronal loops, with the shorter one (loop 1) corresponding to the shorter period. This clearly explains the detection of two periods within one coronal loop system in \cite{Auchere-2018ApJ...853..176A}.

\begin{figure*}
\centering
\includegraphics[width=\textwidth]{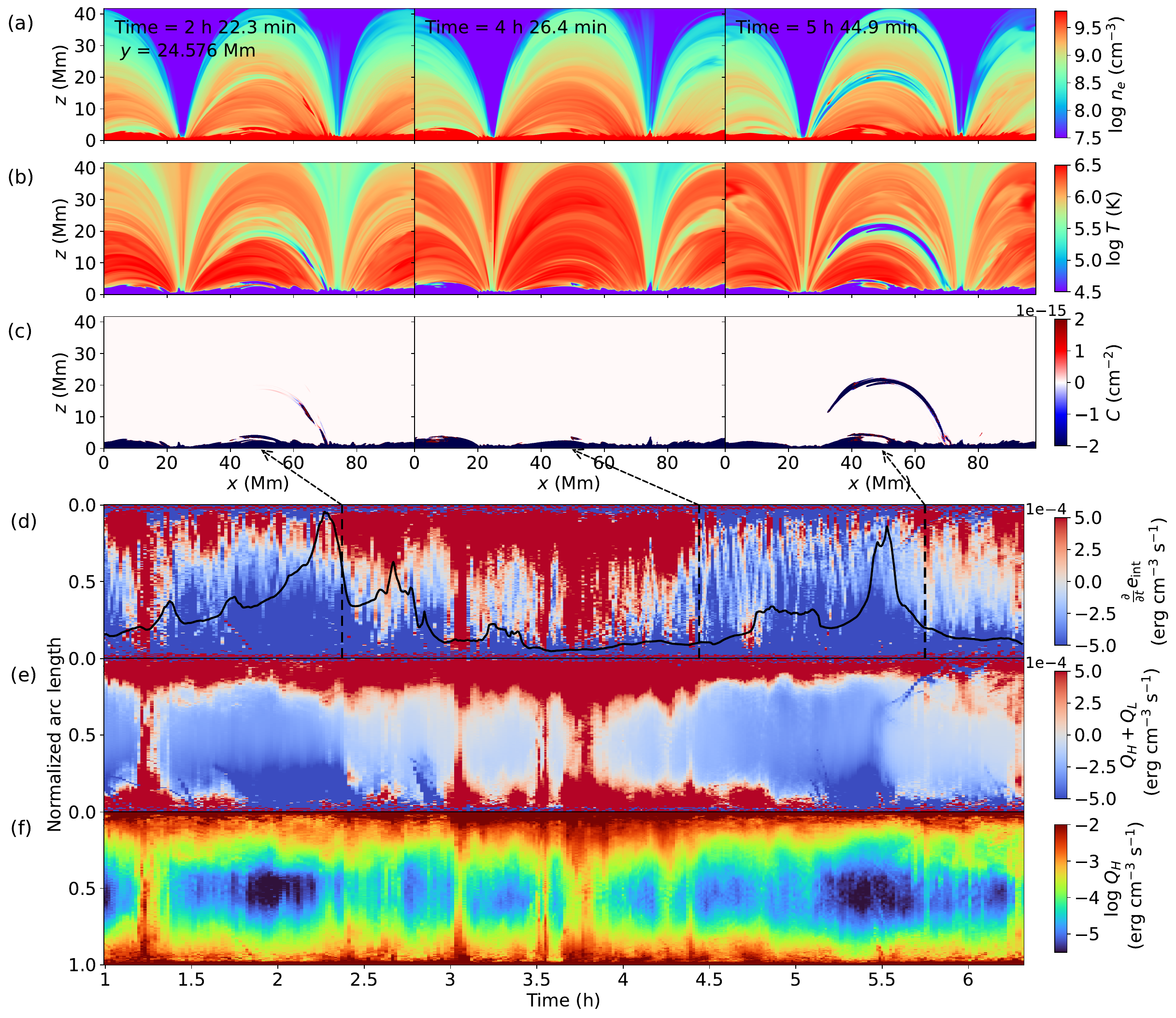} 
\caption{2D distributions of (a) number density, (b) temperature and (c) isochoric TI metric in the $x-z$ plane at three typical time instants. An animation showing the evolution of the quantities depicted in panels (a) -- (c) is available online, covering the time period from $t=1$~h to $t=6$~h~19~min. Temporal evolution of (d) the temporal rate of change of internal energy per unit volume, $\frac{\partial e_{\rm{int}}}{\partial t}$, (e) the sum of volumetric heating and radiative cooling rates, $Q_H+Q_L$, and (f) the volumetric heating rate, $Q_H$, along the loop, based on averaging along 91 magnetic field lines with fixed seed points, as shown in the grey tubes in Figure~\ref{fig:overview}(a). The normalized arc length in panels (d) -- (f) begins at the photosphere of $z=0$~Mm, with corresponding loop length measured 66.25~Mm on time average. The black line in panel (d) is the same as the dash-dot light curve of AIA~304~$\mathrm{\AA}$ in Figure~\ref{fig3:euv_aia}(d).
}
\label{fig4:tne}
\end{figure*}

In essence, such periodic EUV intensity pulsations reflect the evaporation-condensation cycles of the chromosphere-corona system that are under TNE cycles, as indicated by differential emission measure (DEM) analyses in observations \citep{Froment-2015ApJ-DEM...807..158F,Froment-2020A&A...633A..11F}. To illustrate this, we take a temporal segment of the simulation covering the time interval between $t_1$ and $t_2$ as an example for further investigations. First, within the $x-z$ plane passing through the center of the active region, the corona experiences a temperature (mass) cycle of cooling (condensation), heating (evaporation) and cooling (condensation) again at the arched region with an apex height of $\sim20$~Mm (Figure~\ref{fig4:tne}a,b), which corresponds well to the region of coronal loop 1. The isochoric TI-unstable criteria C \citep{Parker-1953ApJ...117..431P,Xia-2011ApJ...737...27X} well reveals the locations where rain blobs forms (Figure~\ref{fig4:tne}c and its online animation),
\begin{equation}
    C \equiv k^2-\frac{1}{\kappa}\left(\frac{\partial Q_H}{\partial T}-\frac{n_e n_H \partial \Lambda(T)}{\partial T}\right)<0.
\end{equation}
Here $k$ represents the wave number of perturbation, which takes the value of twice the condensation scale ($\sim15$~Mm); $\kappa$ is the Spitzer thermal conductivity; $n_e n_H \Lambda(T)$ represents the optically-thin radiative loss. The term ${\partial Q_H}/{\partial T}$ is treated as zero due to the lack of explicit correlation between volumetric heating rate and temperature in this simulation. Furthermore, we define the coronal loop 1 (referred to as ``the loop'' thereafter in Section \ref{sec:results}) in 3D space by integrating magnetic field lines through 91 fixed seed points as exemplified in Figure~\ref{fig:overview}a, and evaluate each term of the energy equation based on Equation~\ref{eq:eint},
\begin{equation}
\label{eq:eint}
\frac{\partial e_{\rm{int}}}{\partial t} =-\frac{\partial (e_{\rm{int}}v_\parallel)}{\partial s}-p\frac{\partial v_\parallel}{\partial s}+\frac{\partial q_{\mathrm{cond}}}{\partial s}+Q_H+Q_L,
\end{equation}
where $e_{\rm{int}}$ denotes the internal energy per unit volume; $q_{\mathrm{cond}}$ represents the conductive heat flux, while $Q_H$ and $Q_L$ represent the volumetric heating and radiative cooling rates, respectively. When $T<2\times10^4$~K and $\tau>10^{-8}$, $Q_L$ equals the solution of radiative transfer equation, which generally is negative in the coronal domain; in other case, $Q_L$ simply equals the optically-thin radiative loss. Note that in Equation \ref{eq:eint}, we describe the loop in 1D by neglecting the expansion or compression perpendicular to the loop. This assumption holds valid within the steady active region, where although there are occasional local motions perpendicular to the magnetic field, the loop overall behaves as a rigid structure.

As shown in Figure~\ref{fig4:tne}d, the internal energy of the coronal plasma undergoes a cyclic evolution of decreasing, increasing and then decreasing. The decreasing (increasing) internal energy leads to cooling (heating) of corona, which is in phase with AIA 304~$\rm{\AA}$ lightcurve of the loop. At the end of each cooling process, coronal rains form dramatically, along with the 304~$\rm{\AA}$ lightcurve reaching its peak. This clearly proves that both periodic coronal rains and periodic EUV pulsations are manifestations of periodic energy evolution of coronal loops under TNE state. More importantly, since the 1D loop quantity in Figure~\ref{fig4:tne}d represents the quantities averaged among 91 magnetic field lines with a loop apex cross section of $3\times6\ \rm{Mm^2}$ (Figure~\ref{fig:overview}a; a typical loop scale as reported in \citealt{Aschwanden-2013SoPh..283....5A}), our results reveal in 3D that the neighboring coronal strands evolve in phase. This explains why coronal rain showers form and fall synchronously and in the immediate area \citep{Antolin-2012ApJ-fine-rains...745..152A,Froment-2020A&A...633A..11F,Antolin-2023A&A...676A.112A}. Additionally, sum of the two energy source terms, $Q_H+Q_L$, manifests a largely similar evolution pattern as that of sum of all energy terms (Figure~\ref{fig4:tne}e). This suggests that focusing solely on heating and cooling may suffice for a rough evaluation of the long-term energy evolution of coronal loops.

\subsection{Self-consistent heating rates} \label{subsec:heating rates}

\begin{figure}
\centering
\includegraphics[width=0.485\textwidth]{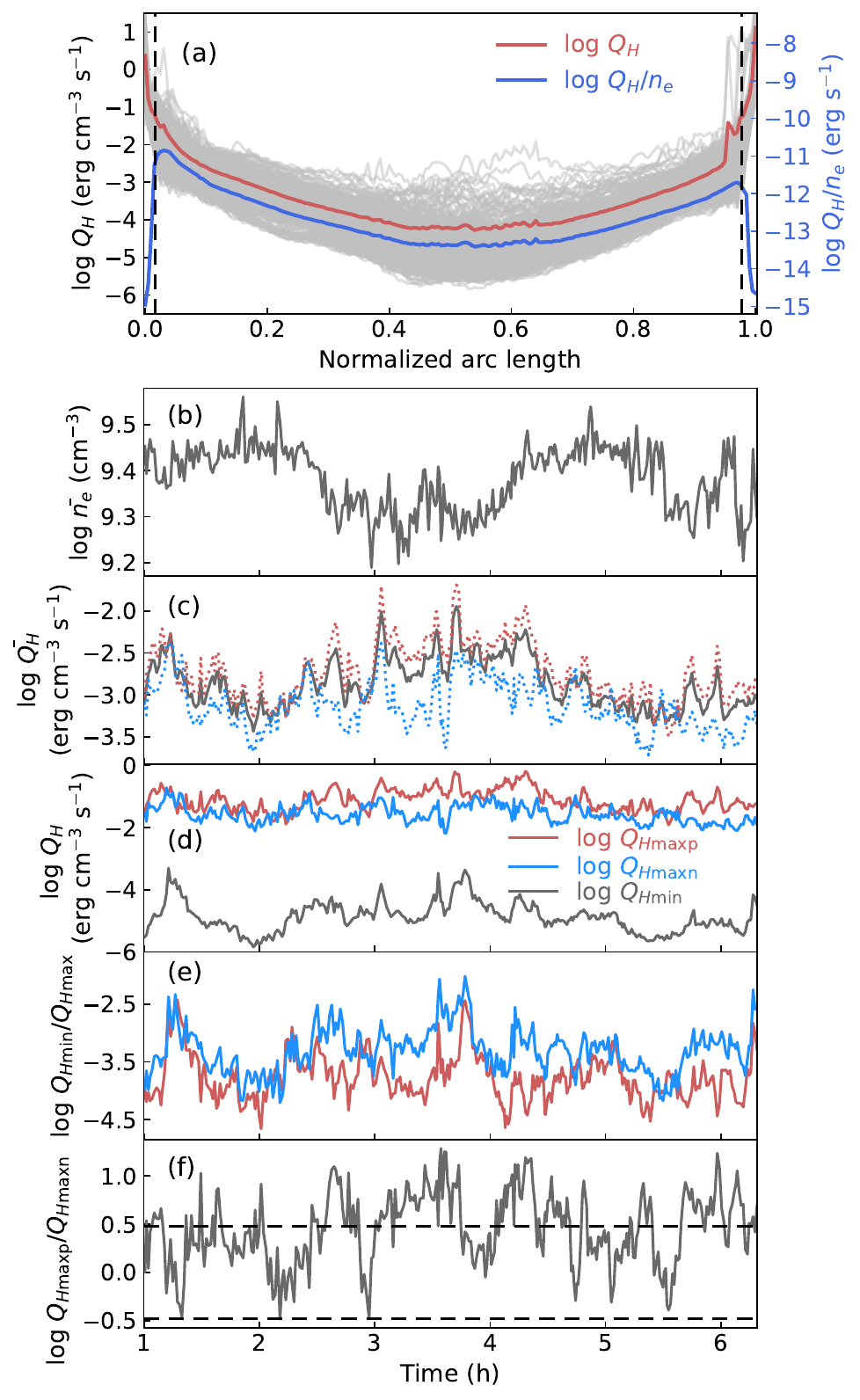} 
\caption{(a) Distributions of the volumetric heating rate, $Q_H$, along the loop at all time steps (grey lines) during the evolution as in Figure~\ref{fig4:tne}(f), where the coronal loop is defined by its portion with densities below $10^{-12}\ \mathrm{g}\ \mathrm{cm^{-3}}$. The red and blue lines represent the time-averaged profile of $Q_H$ and $Q_H/n_e$, respectively. The black dashed lines denote the time-averaged position of footpoints of the coronal loop. Panels (b) and (c) display temporal evolution of the number density, $\bar{n}_e$, and volumetric heating rate, $\bar{Q}_H$, that are averaged along the coronal loop. The red and blue lines denote the values averaged within the half coronal loop close to positive and negative polarities, respectively. Panel (d) shows $Q_H$ at positive footpoint (in red), negative footpoint (in blue) and apex (in dark-grey), as defined by the maxima around footpoints and the minimum over the loop. Panels (e) and (f) display the apex to footpoint heating ratio and footpoints heating asymmetry. The black dashed lines in panel (f) denote the asymmetry values of 3 and 1/3.}
\label{fig5:heating_rates}
\end{figure}

The coronal heating process continuously drives the periodic mass circulation of the chromosphere-corona system. Intuitively, Figure~\ref{fig4:tne}f presents spatial stratification and temporal evolution of the volumetric heating rate, accompanied by different levels of impulsive heating events. Such features closely align with those found in previous MURaM-based active region simulations \citep{Rempel-2017ApJ...834...10R,ChenF-2022ApJ...937...91C}. Nevertheless, previous studies primarily focused on general heating properties across the entire active region, while this work reveals more details of heating characteristics along a single coronal loop.

Figure~\ref{fig5:heating_rates}a depicts volumetric heating rate profiles along the loop at all time steps in grey, which indicate a clear decreasing trend from the base to the apex. The time-averaged heating profile measures the decrease from around 1 to $10^{-4}$~erg~cm$^{-3}$~s$^{-1}$. Remarkably, such a decreasing trend along the loop is complex, with a higher-order decrease rate in the chromosphere and lower coronal regions (normalized arc length $<0.05$) than that in its main body. Moreover, considering the large density difference between chromosphere and corona, the normalized heating quantity $Q_H/n_e$, representing the heating rate per particle, serves as a more direct measure of the heating-induced temperature change rate. It reaches maximum in the transition region and then decreases with height in a similar trend, with maximum values of $1.41\times10^{-11}$~erg~s$^{-1}$ near the positive polarity and $1.98\times10^{-12}$~erg~s$^{-1}$ near the negative polarity. According to the definition, footpoint of coronal loop is located in the transition region, which in our simulation corresponds to the first pixel with density below $10^{-12}$~g~cm$^{-3}$. Subsequent calculations are conducted within the regions bounded by the two footpoints.

Along a coronal loop, the thermodynamic evolution of the plasma is highly controlled by heating rates, variations in which lead to divergent evolutionary paths \citep{Froment-2018ApJ-1d-tne...855...52F,Klimchuck-2019ApJ-tne...884...68K,Johnston-2019A&A...625A.149J,Pelouze-2022A&A...658A..71P}. Consequently, it is important to further assess the self-consistent heating rate through a number of characteristic quantities, including its magnitude, timescale, stratification, and asymmetry of the heating rate. 
\begin{enumerate}
\itemsep0em
\item Above all, as a consequence of the heating process, the spatially-averaged number density exhibits cyclic evolution ranging from around $10^{9.2}$ to $10^{9.5}$~cm$^{-3}$ (Figure~\ref{fig5:heating_rates}b).
\item Instead of being quasi-static over time, the spatially averaged volumetric heating rate of the loop, $\bar{Q}_H$, observably varies with time (Figure~\ref{fig5:heating_rates}c). On the timescale of a TNE cycle, it demonstrates a gradual variation of around one order of magnitude, within the range of $10^{-3.2}$ to $10^{-2.2}$~erg~cm$^{-3}$~s$^{-1}$. On a timescale of a few minutes, it manifests as a series of impulsive heating events, among which the most intensive one ($t \sim3$~h) explosively increases by about ten times.
\item The heating rate maintains a significant stratification over evolution. As shown in Figure~\ref{fig5:heating_rates}d,e, the minimum-to-maximum (apex-to-footpoint) ratio generally fluctuates around $10^{-3.5}$, at most ranging from around $10^{-4.5}$ to $10^{-2.5}$. Such level of heating stratification well accords with the rule of thumb of 0.1 for TNE state \citep{Klimchuck-2019ApJ-tne...884...68K}.
\item The heating rate holds a strong and variant asymmetry over time (Figure~\ref{fig5:heating_rates}f). First, the heating rate on the positive polarity side is notably higher than that on the negative side (see also red and blue curves in Figure~\ref{fig5:heating_rates}c,d), which naturally arises from the asymmetry of the magnetic field strength (Figure~\ref{fig:overview}c) and explains why most coronal rains fall toward the negative polarity (Figure~\ref{fig3:euv_aia}c). More importantly, over time, the heating asymmetry exhibits significant variability across different timescales, with values ranging from -1/3 to around 10 (Figure~\ref{fig:overview}f). Such an asymmetry exceeds the threshold of the TNE condition given by \cite{Klimchuck-2019ApJ-tne...884...68K}, i.e. -1/3 to 3, during about half of the time.
\end{enumerate}

\section{Summary and Discussions} \label{sec:conclusion}

In this work, we present the first 3D RMHD modelling of an active region with periodic coronal rains, concurrent with periodic EUV intensity pulsations. By including a convective zone extending to around 7.5 Mm in depth, the magneto-convection drives heating processes self-consistently. As a result, on large scales, periodic chromosphere-corona mass circulations occur within different coronal loops. On small scales, coronal rains form as synchronous and neighbouring rain showers with unprecedented multi-thermal fine structures, especially the dense cores in chromospheric temperatures, as well as the thin and sharp envelope in transition region temperatures. On large scales, coronal rains recur within two coronal loops with periods of around 3.0~h and 3.7~h, accompanied with periodic EUV intensity pulsations across multiple wavelengths. Moreover, as closely linked to the coronal heating problem, we quantitatively reveal that the coronal heating rate behind the periodic coronal rains is highly stratified, fluctuating and asymmetric.

How the near-surface magneto-convection drives energy releases along coronal loops has been investigated mainly in 3D RMHD quiet Sun simulations \citep{Kohutova-2020A&A...639A..20K,Breu-2022A&A...658A..45B,Kuniyoshi-2023ApJ...949....8K}. By comparison, our model shares one characteristic with them that the coronal heating rate is highly bursty along loops, which may represent a common feature of coronal heating processes driven by the turbulent convective motions \citep{Martinez-Pillet-2013SSRv..178..141M,Rempel-2014ApJ...789..132R}. Due to the largely independent magneto-convection environments at the footpoints of coronal loops, such dynamic and bursty heating rates naturally create asymmetry in heating on both sides. This heating asymmetry serves as an inhibiting factor for TNE \citep{Froment-2018ApJ-1d-tne...855...52F,Klimchuck-2019ApJ-tne...884...68K,Pelouze-2022A&A...658A..71P}. Nevertheless, as an active region simulation with a magnetic field strength of $>2$~kG, our model differs from them in two aspects. First, in addition to these minute-level bursts, the coronal heating rate shows long-term variations in hours, which reflects the intrinsic evolution of the active region. Second, importantly, our model maintains higher coronal heating flux (and stronger heating stratification) than that in quiet Sun models, which facilitates more tolerant TNE conditions \citep{Froment-2018ApJ-1d-tne...855...52F,Klimchuck-2019ApJ-tne...884...68K,Pelouze-2022A&A...658A..71P} and explains why a similar heating asymmetry (reaching ten times) leads to different evolutionary tracks, namely one-off coronal rains in \cite{Kohutova-2020A&A...639A..20K} while periodic ones in our model. In this sense, the periodic coronal rains would be more likely to occur in active regions.

Among 3D RMHD active region simulations, our model describes a steady bipolar active region, which is mainly driven by magneto-convection and represents a common magnetic configuration on the Sun. This differs from models that include additional drivers, such as flux emergence or interactions between different sunspots, which may lead to even stronger or more impulsive energy releases \citep{Cheung-2019NatAs...3..160C, ChenF-2022ApJ...937...91C,Toriumi-2023NatSR..13.8994T, Rempel-2023ApJ...955..105R,Wang-2023ApJ...956..106W}. For example, in an emerging active region with a fan-spine-like magnetic topology, continuous flux emergence can keep driving magnetic reconnections at a current sheet above coronal loops, providing a persistent heating flux that is one order of magnitude larger than that along coronal loops \citep{Lu-2024NatAs...8..706L}.

Fundamentally, evolution of coronal plasma attributes to the coronal heating, which is often added to the energy equation as a user-defined source term. In this context, our model allows to depict distribution and evolution of the coronal heating rate self-consistently driven by magneto-convection, which enhances our understanding of the coronal heating process from a quantitative perspective. Temporally, it is notably more fluctuating compared to commonly-used static heating profiles \citep{Mikic-2013ApJ...773...94M,Fang-2013ApJ...771L..29F,Fang-2015ApJ...807..142F,Xia-2017A&A-rain...603A..42X,Froment-2018ApJ-1d-tne...855...52F,Pelouze-2022A&A...658A..71P}, and more complex than repeatedly-impulsive ones \citep{Johnston-2019A&A...625A.149J}. In this sense, the randomized heating function derived from a power-law heating spectrum \citep{Zhouyh-2020NatAs...4..994Z,Lixh-2022ApJ-rain...926..216L} largely captures the dynamic and complex characteristics. Spatially, the heating rate decrease piecewise with respect to arc length $s$ (Figure~\ref{fig5:heating_rates}a) and height $z$ (Figure~\ref{efig:heating_height}), with three different rates of decline at different arc length or height. This is different from the simple heating function $\propto \exp{(-s/\lambda)}$ used in 1D simulations \citep{Mikic-2013ApJ...773...94M,Froment-2018ApJ-1d-tne...855...52F,Pelouze-2022A&A...658A..71P} and that $\propto \exp{(-(z/\lambda)^2)}$ used in 2.5D \citep{Fang-2013ApJ...771L..29F,Fang-2015ApJ...807..142F,Lixh-2022ApJ-rain...926..216L} and 3D \citep{Xia-2017A&A-rain...603A..42X} simulations ($\lambda$ represents the heating scale height). In conclusion, such a self-consistent heating rate exhibits considerable levels of complexity in time and space. We have made the heating rate with respect to height and normalized arc length at every time steps publicly available in a Zenodo repository (doi:\href{https://doi.org/10.5281/zenodo.13308624}{10.5281/zenodo.13308624}), which is expected to provide a theoretical reference and contribute to the definition of heating functions in a wide range of relevant studies, including coronal rains, coronal loops \citep{Klimchuk-2010ApJ...714.1239K,Lionello-2013ApJ...773..134L,Dai-2024ApJ...965....2D} and prominence \citep{Xia-2016ApJ...823...22X,Guo-2021ApJ...920..131G,Ni-2022A&A...663A..31N} on the Sun and stars \citep{Daley-Yates-2023MNRAS.526.1646D}.

\begin{acknowledgments}
We thank Jun Chen, Chen Xing, Yu Dai, Jie Hong, Yuhao Zhou, Rony Keppens and James A. Klimchuk for their valuable discussions. This work is supported by National Key R\&D Program of China under grants 2021YFA1600504 and NSFC under grants 12373054. Z.L. is supported by the Postgraduate Research \& Practice Innovation Program of Jiangsu Province KYCX22\_0107. C.W. is funded by the Postgraduate Research \& Practice Innovation Program of Jiangsu Province KYCX23\_0118.
\end{acknowledgments}

\appendix
\section{Evolution of the simulation with refined mesh}
In this work, we interpolate a simulation snapshot from \cite{Rempel-2017ApJ...834...10R} with nearest neighbour method from $192\times192\times64$~km$^{3}$ to $96\times96\times32$~km$^{3}$ as the initial state, which led to unphysical oscillations of MHD quantities at first and took about half a solar hour to reach a new dynamic equilibrium (Figure \ref{efig:relaxation}). The new relaxed state moderately deviates from the initial state due to the evolution of coronal plasma under TNE and the changes brought in by the mesh refinement.
\setcounter{figure}{0}
\renewcommand\thefigure{A\arabic{figure}}
\begin{figure}
    \centering
    \includegraphics[width=0.485\textwidth]{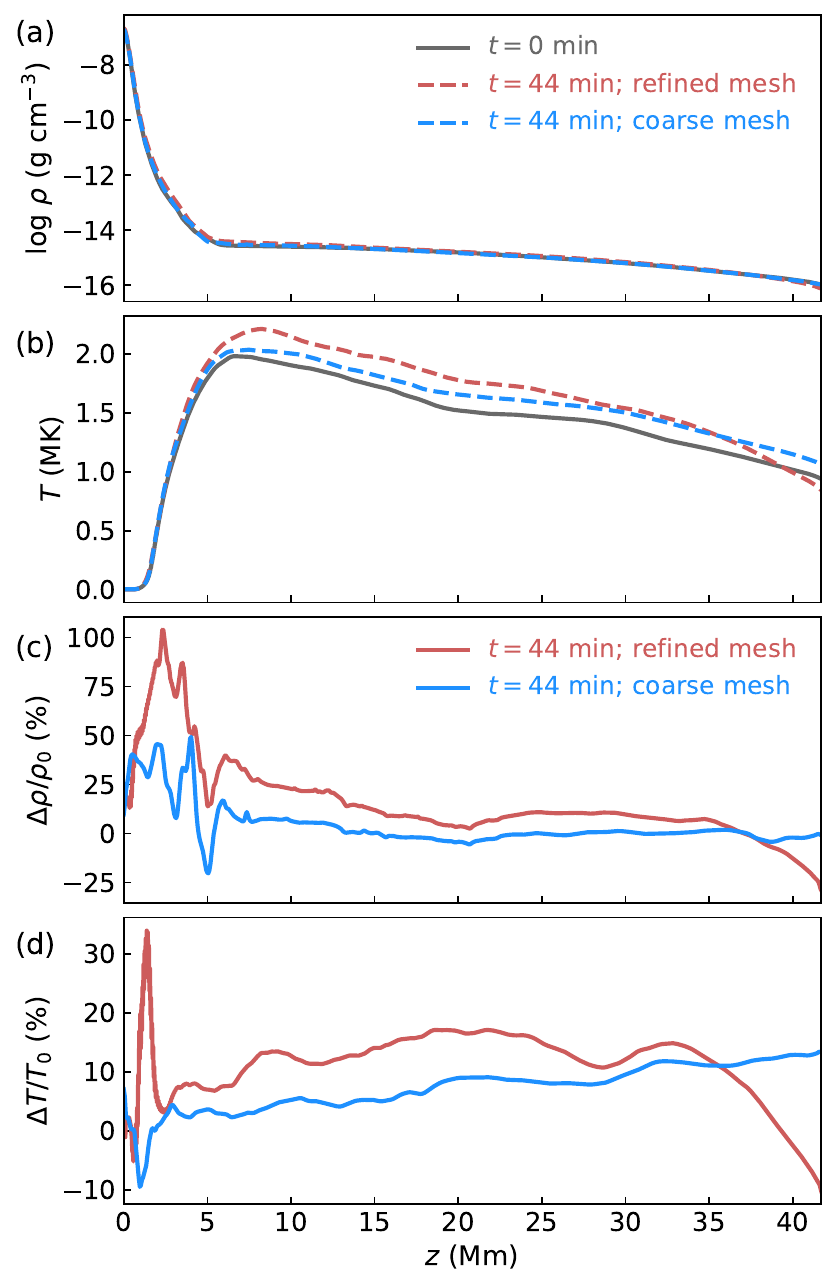}
    \caption{The horizontally averaged height-dependent profiles of (a) mass density and (b) temperature at $t=0$~min and $t=44$~min. The percentual differences of the quantities with respect to the initial state are shown in panels (c) and (d). The red and blue lines represent the profile of simulation snapshot with the refined mesh ($96\times96\times32$~km$^{3}$) and coarse mesh ($192\times192\times64$~km$^{3}$), respectively.
    }
    \label{efig:relaxation}
\end{figure}

\section{$\rm{H \alpha}$ imagae synthesis}
Assuming that the $\rm{H \alpha}$ line source function of the coronal rain is dominated by the scattering of the incident radiation from the lower atmosphere, we can simplify the solution of the radiative transfer as
    \begin{equation}
    I(\nu)=S(1-e^{-\tau_{\nu}}
),
    \label{eq:transfer1}
    \end{equation}
where $S$ represents the $\rm{H \alpha}$ line source function, which is found almost constant at the same altitudes and we adopt a value of $2.98\times10^{-6}\ \rm{erg}\ \rm{cm}^{-2}\ \rm{sr}^{-1}\ \rm{Hz}^{-1}$ for the altitude of $\sim20$~Mm, as reported in \cite{Heinzel-2015A&A...579A..16H}; $\tau_{\nu}$ represents the optical depth that is defined along line of sight (the $y$-direction in Figure~\ref{fig2:rain_slcie}d) by
    \begin{equation}
    \mathrm{d}\tau_{\nu}(y) = \kappa_\nu(y)\mathrm{d} y = 1.7\times10^{-2}n_2(y)\phi_\nu(y)\mathrm{d} y,
    \label{eq:transfer2}
    \end{equation}
where $n_2$ denotes the hydrogen second-level population; $\kappa_\nu$ denotes the absorption coefficient for the $\rm{H\alpha}$ line; $\phi_\nu$ represents the normalized absorption profile as defined in \cite{Heinzel-2015A&A...579A..16H}. To determine $n_2$, we assume that $n_2/n_{\rm{H}}$ and the temperature $T$ are in one-to-one correspondence and obtain such relation from the atmospheric data of RADYN simulation \citep{Carlsson-2012A&A...539A..39C}. Therefore, by first integrating the optical depth in Equation \ref{eq:transfer2} and then substituting it into Equation \ref{eq:transfer1}, we are able to calculate the specific intensity at the $\rm{H\alpha}$ line center shown in Figure~\ref{fig2:rain_slcie}d.

\section{Periods of EUV intensity pulsations}

In Figure~\ref{fig3:euv_aia}, we display five peak times of synthetic AIA 171 $\mathrm{\AA}$ light curve and attribute them to periodic intensity pulsations of two coronal loops (loop 1: $t_1$, $t_2$, $t_4$; loop 2: $t_1$, $t_3$, $t_5$). As a complement, Table \ref{tab:period} provides detailed peak times and the calculations of time differences, as well as periods 
of pulsations for each loop.
\renewcommand\thetable{A\arabic{table}}
\begin{table}[ht]
\tabletypesize{\scriptsize}
    \centering
    \begin{tabular}{c c c c}
        \hline\hline
        & AIA 171 $\mathrm{\AA}$ & Loop 1 & Loop 2 \\
        & peak time & $\Delta t$ (h) & $\Delta t$ (h) \\
        \hline
       \multirow{5}{*}{Time}  & 2 h 22.6 min ($t_1$) & & \\
         & 5 h 13.6 min ($t_2$) & 2.85 ($t_2-t_1$) & \\
         & 6 h 11.1 min ($t_3$) &  & 3.81 ($t_3-t_1$)\\
         & 8 h 15.5 min ($t_4$) & 3.03 ($t_4-t_2$) & \\
         & 9 h 48.6 min ($t_5$) &  & 3.62 ($t_5-t_3$) \\
         \hline
        Average &  & 2.94 & 3.72 \\
        \hline
    \end{tabular}
    \caption{Peak times of the synthetic AIA 171 $\mathrm{\AA}$ light curve and the time differences between the consecutive peaks in two coronal loops, $\Delta t$. Notations within brackets represent peak times from $t_1$ to $t_5$ and the two peaks involved in the time difference calculation.}
    \label{tab:period}
\end{table}

\section{The volumetric heating rate with respect to height}

In 2.5D or 3D simulations, the heating rate is usually written as a function of height. Therefore, for readers' reference, we also plot the volumetric heating rate (Figure~\ref{fig5:heating_rates}a) with respect to height in Figure~\ref{efig:heating_height}, where panels (a) and (b) depict the heating rate of the half loop near positive and negative polarity, respectively. The data and the script to plot Figure~\ref{efig:heating_height} are available in a Zenodo repository (doi:\href{https://doi.org/10.5281/zenodo.13308624}{10.5281/zenodo.13308624}).

\renewcommand\thefigure{A\arabic{figure}}
\begin{figure}
    \centering
    \includegraphics[width=0.485\textwidth]{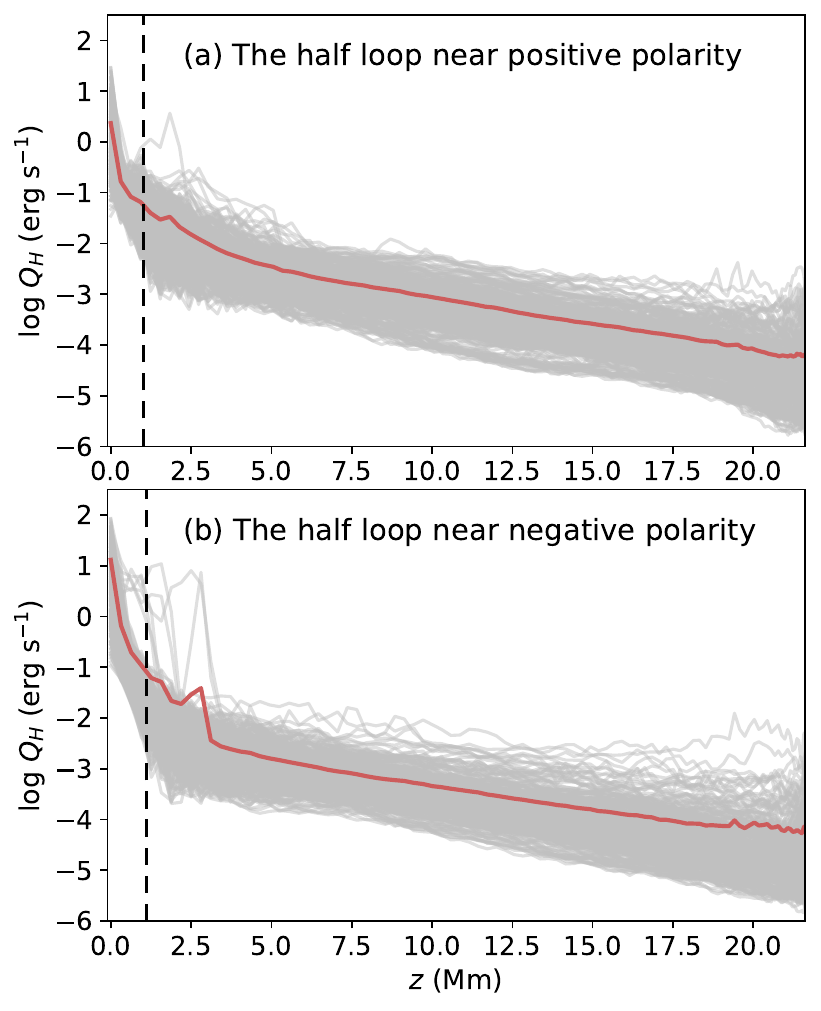}
    \caption{Distributions of the volumetric heating rate, $Q_H$, with respect to height, $z$, at all time steps (grey lines) during the evolution as in Figure~\ref{fig4:tne}(f). Panels (a) and (b) depict the heating rate of the half loop near positive and negative polarity, respectively. The red line represents the time-averaged profile of $Q_H$. The coronal loop is defined by its portion with densities below $10^{-12}\ \mathrm{g}\ \mathrm{cm^{-3}}$. The black dashed lines denote the time-averaged position of footpoints of the coronal loop.}
    \label{efig:heating_height}
\end{figure}

\bibliography{zbibref1}{}
\bibliographystyle{aasjournal}



\end{document}